\documentclass[%
 reprint,
 amsmath,amssymb,
 aps,
]{revtex4-2}

\usepackage{graphicx}
\usepackage{dcolumn}
\usepackage{tabularx,booktabs}

\providecommand{\apj}{ApJ}
\providecommand{\jcap}{JCAP}

\usepackage{bm}

\begin{document}

\preprint{APS/123-QED}

\title{Flux of Extragalactic Dark Matter in Direct Detection Experiments}

\author{Shokhruz Kakharov}
 \email{shokhruzbekkakharov@college.harvard.edu}
 \altaffiliation[Also at ]{Physics Department, Harvard University.}
\author{Abraham Loeb}%
 \email{aloeb@cfa.harvard.edu}
\affiliation{%
 Astronomy Department, Harvard University,\\
 60 Garden St., Cambridge, MA 02138, USA
}%

\date{\today}

\begin{abstract}
We calculate the contribution of extragalactic dark matter to the local dark matter density and flux in the Milky Way. By analyzing the Galactic escape velocity as a function of direction, we establish a criterion for separating particles bound to the Milky Way from those originating in the Local Group environment. Our analysis finds that about 26\% of local particles are unbound to the Galaxy and contribute about 38\% of the total mass flux. The sky pattern is anisotropic. We provide rate and modulation predictions relevant for current and future direct-detection experiments.
\end{abstract}

\maketitle


\section{Introduction}

Dark matter constitutes about 85\% of the matter content in our universe, yet its nature remains elusive. Past predictions for the detection rate of laboratory experiments focused on particles that are gravitationally-bound to the Milky Way~\cite{Lewin1996,Jungman1996,Bertone2005,Freese2013}.Here, we show that a significant component originates from beyond our Galaxy. Inclusion of the extragalactic contribution could impact the quantitative interpretation of direct detection experiments.~\cite{2021PhLB..82036551H}

To identify extragalactic dark matter particles, we calculate the escape velocity as a function of direction across the sky. Particles exceeding this threshold cannot be gravitationally bound to the Milky Way and must have an extragalactic origin. This condition provides a method for distinguishing galactic from extragalactic particles; those moving faster than the local escape velocity in a given direction must originate beyond the Milky Way's gravitational influence~\cite{2013ApJ...771..117B}

The extragalactic component primarily originates from the Local Group environment surrounding our Galaxy. The Local Group, dominated by the Milky Way, Andromeda and the intergalactic matter around them, contains of order $\sim 5\times 10^{12}M_{\odot}$ of mass, with the baryonic component representing approximately 17\% of the total~\cite{Rubin2014}. This intergalactic medium supplies high-velocity dark matter particles, potentially detectable in Earth-based experiments. Recent N-body simulations~\cite{DeBrae2025} have demonstrated that the highest velocity dark matter particles (v $>$ 600 km/s) in Milky Way-like halos predominantly originate from fast-moving substructure, particularly Large Magellanic Cloud analogs, rather than from the broader extragalactic environment. Hydrodynamical Local-Group simulations independently confirm this picture and find no evidence for a separate extragalactic high-speed component, with the local velocity distribution well described by a truncated Maxwellian tail~\cite{2024JCAP...03..046S}.This finding suggests a multi-component environment of high-velocity dark matter sources that must be considered in detection experiments.

The angular dependence of the galactic escape velocity results from the non-spherical nature of the Milky Way's gravitational potential and from our motion through the halo. Directions where the threshold for extragalactic particles is lower, enhance their detection potential.

In \S II, we calculate the Galactic escape velocity distribution and find the Local Group dark matter contribution in comparison to the Milky Way's dark matter. Subsequently, we explore the implications for direct detection experiments with quantitative predictions for observable signatures.

\section{Galactic Escape Velocity}
\label{sec:methodology}

We model the escape velocities in a Galactocentric coordinate system where the x-axis points from the Galactic center toward the Sun, the y-axis points in the direction of Galactic rotation, and the z-axis points toward the North Galactic Pole, forming a right-handed system. In this framework, the Local Standard of Rest (LSR) moves primarily along the positive y-axis with velocity $\vec{v}_{\text{LSR}} = (0, 240, 0)$ km/s\cite{Schoenrich2010}.

For our directional analysis, we use spherical coordinates where:
\begin{itemize}
    \item $\theta$ is the polar angle measured from the positive z-axis (North Galactic Pole), with $\theta \in [0, \pi]$
    \item $\phi$ is the azimuthal angle measured in the x-y plane from the positive x-axis (toward the Sun) in the counterclockwise direction, with $\phi \in [0, 2\pi]$
\end{itemize}

In these coordinates, the LSR direction corresponds to $\theta = \pi/2$ (Galactic plane) and $\phi = \pi/2$ (90° from the Sun-Galactic center line).

We first model the escape velocities with a dipole term based on the Local Standard of Rest (LSR) velocity \cite{Schoenrich2010}. Escape velocities are calculated using orbital integration with the McMillan (2017) Galactic potential~\cite{McMillan2017} implemented through the GalPot software library~\cite{McMillan2017}.

Subsequently, we fit the residuals using quadrupole moments derived from spherical harmonics. For the quadrupole analysis, we implement the following spherical harmonics:

\begin{itemize}
    \item $Y_{2,0}(\theta) = 0.5(3\cos^2\theta - 1)$ represents prolate (positive values) or oblate (negative values) deformation along the z-axis. It is independent of the azimuthal angle $\phi$ and shows deformation that preserves axial symmetry around the Galactic pole axis.
    
    \item $Y_{2,1}(\theta,\phi) = \sin(\theta)\cos(\theta)\cos(\phi)$ represents asymmetry along a tilted axis, breaking the reflection symmetry across the Galactic plane (xy-plane). This component can indicate asymmetric features between the northern and southern Galactic hemispheres.
    
    \item $Y_{2,2}(\theta,\phi) = \sin^2(\theta)\cos(2\phi)$ represents ellipticity in the Galactic plane (x-y plane), showing a 4-fold pattern with two positive and two negative regions. This component can identify deviations from axisymmetry in the Galactic disk, such as bar or spiral arm influences.
\end{itemize}

The unit vector pointing in direction $(\theta,\phi)$ is given by:
\begin{equation}
\hat{d}(\theta,\phi) = 
\begin{pmatrix}
\sin\theta \cos\phi \\
\sin\theta \sin\phi \\
\cos\theta
\end{pmatrix}
\end{equation}

This coordinate system allows us to systematically analyze the directional dependence of escape velocities and relate the observed patterns to physical features of the Galactic potential.

Our theoretical model of the smooth potential of the Milky-Way galaxy GalPot model~\cite{McMillan2017} has a significant ellipticity in the x-y plane (strong $Y_{2,2}$ term), some flattening along the z-axis (negative $Y_{2,0}$ coefficient), and the elliptical component is oriented at -90° in the galactic plane.

Our best-fit values are an average escape velocity of 547 km/s, an LSR velocity of 240 km/s, a $Y_{2,0}$ amplitude of -18.3 km/s, a $Y_{2,2}$ amplitude of 27.4 km/s (phase: -90°), and negligible $Y_{2,1}$ terms (of order 0.01 km/s).

While the dipole pattern from the LSR motion explains 98.8\% of the variance, the quadrupole terms account for virtually all remaining variance in the model. Our escape velocity threshold of approximately 520 km/s for identifying extragalactic particles is consistent with recent simulations~\cite{DeBrae2025}, which identify particles exceeding ~600 km/s as predominantly originating from substructure rather than smooth accretion, supporting our criterion for distinguishing galactic and extragalactic components.

\begin{figure}[htbp]
    \centering
    \includegraphics[width=250px]{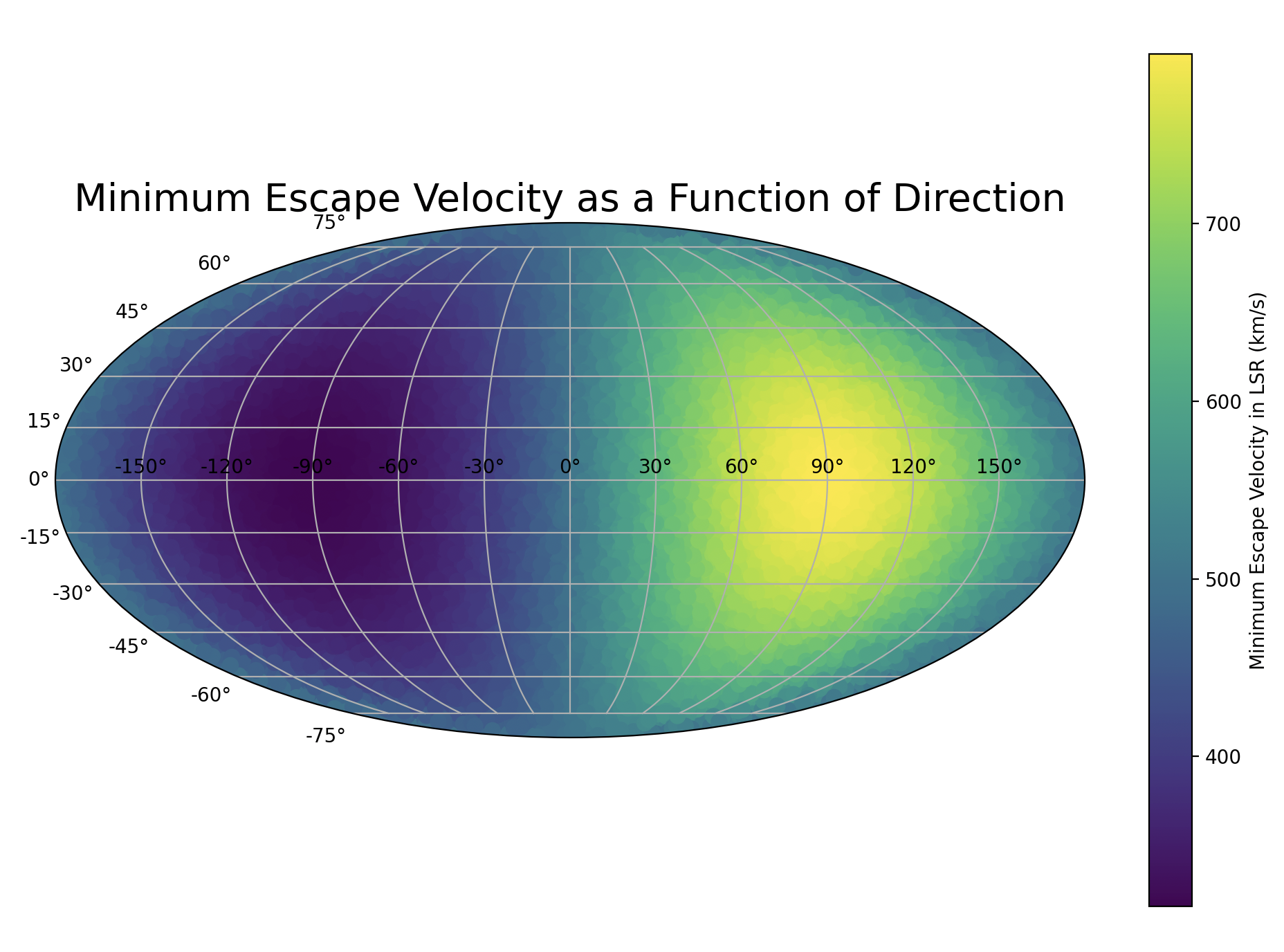}
    \caption{Minimum escape velocity as a function of direction in a spherical projection of Galactic coordinates within our model. The map displays significant angular variation, with the highest escape velocities ($>700$ km/s, yellow region) concentrated around galactic coordinates $(90^{\circ}, 0^{\circ})$ in the eastern hemisphere, while the lowest escape velocities ($<400$ km/s, dark purple) appear in the opposite direction. This anisotropy reflects the LSR motion and the asymmetric gravitational potential of the Milky Way.}
    \label{fig:min_escape_velocity}
\end{figure}

\section{Extragalactic Dark Matter Flux}
\label{sec:darkmatter}
We adopt the standard radial density profile for dark matter in the Local Group~\cite{Navarro1997,Klypin2002,Rubin2014}:
\begin{equation}
\rho(R) = \rho_0 / [(R/R_{\text{vir}}) \cdot (1 + c \cdot R/R_{\text{vir}})^2]
\end{equation}



We define "extragalactic" dark matter as particles originating from the broader Local Group environment outside the Milky Way's virial radius. This primarily consists of the medium surrounding the Milky Way within the Local Group, including the contribution from Andromeda (M31). 

\begin{figure}[htbp]
    \centering 
    \includegraphics[width=250px]{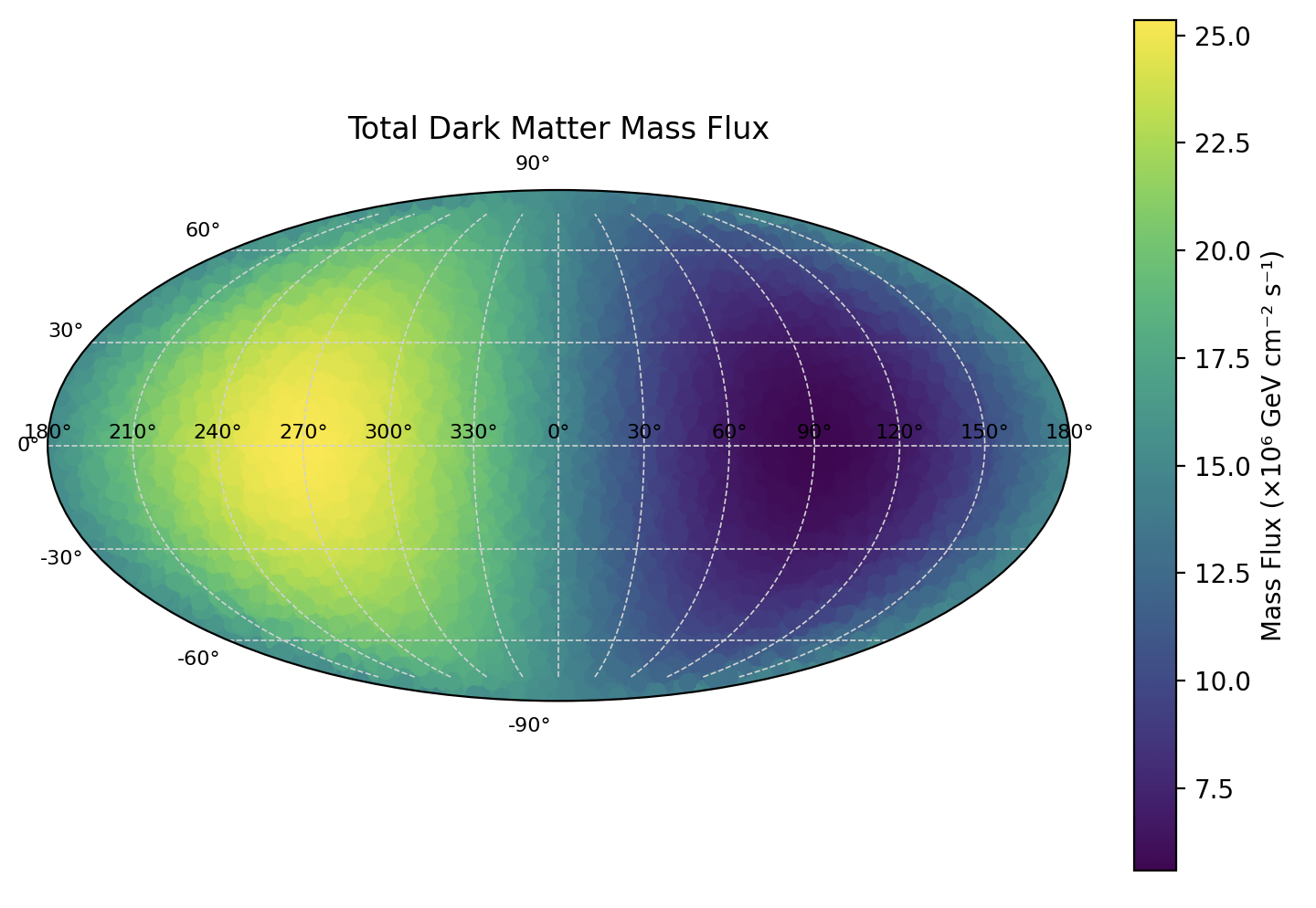}
    \caption{Total mass flux of dark matter ($\rho {\bf v}$) in GeV/cm$^2$/s. The map in galactic coordinates shows a higher flux (yellow) in the western galactic hemisphere.}
    \label{fig:dark_matter_flux}
\end{figure}


Our model of the Local Group explicitly incorporates both the Milky Way and Andromeda halos, along with the diffuse dark matter distribution between them. The combined density exceeds the Milky Way contribution by approximately 10\%, indicating that extragalactic sources make a non-negligible contribution to the local dark matter density. This is consistent with theoretical expectations for the Local Group medium, which is predicted to contain substantial diffuse mass distributed in an NFW profile centered on the Local Group center of mass~\cite{Rubin2014}.

The velocity dispersion of dark matter particles represents a critical parameter in detection experiments. Our comparison between isotropic and anisotropic dispersion models revealed significant differences in the predicted detection rates. The anisotropic model, with its reduced vertical dispersion (170 km/s compared to 270 km/s in the isotropic case), predicts a lower flux from high galactic latitudes, modifying the expected annual modulation signal by approximately 15\%. 

\begin{figure}[htbp]
    \centering
    \includegraphics[width=250px]{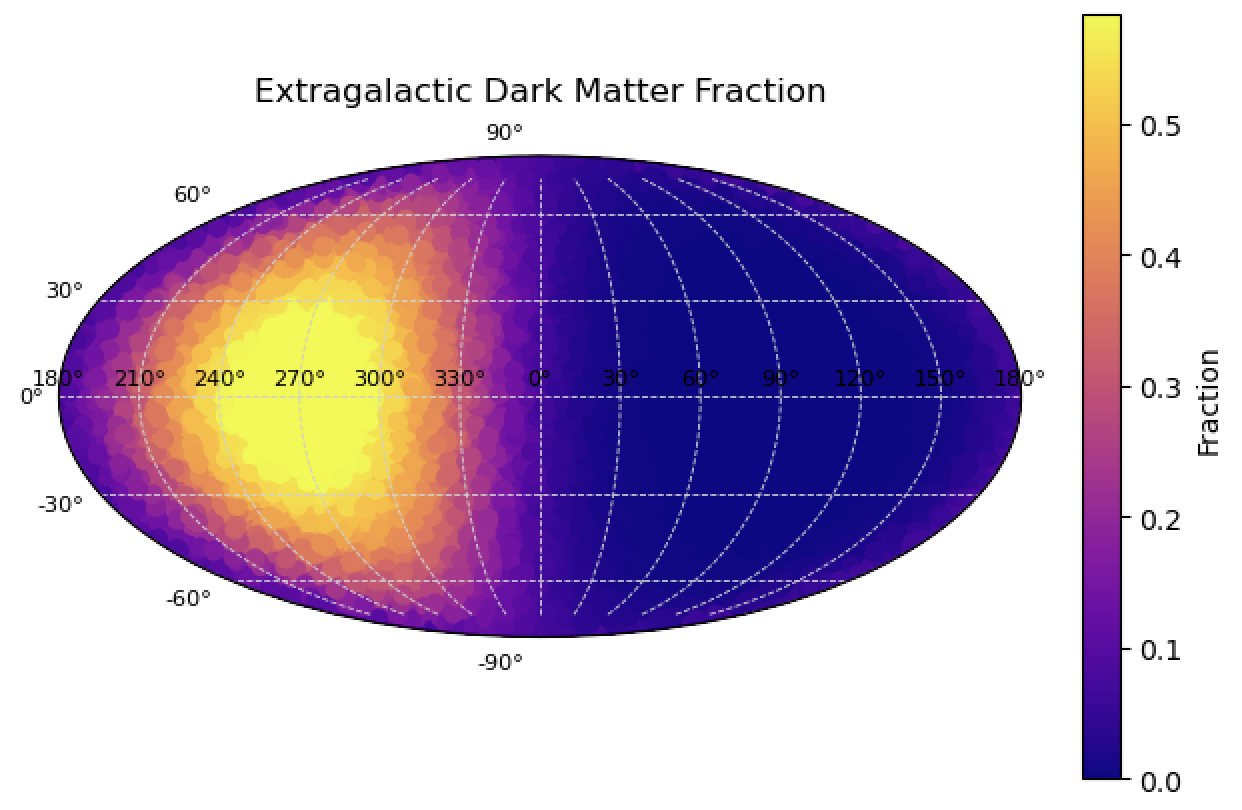}
    \caption{The extragalactic dark matter fraction shows a strong dipole asymmetry with significantly higher extragalactic dark matter contribution (yellow/orange) of up to 70\% in the western galactic hemisphere, contrasted with minimal contribution (dark purple) in the eastern hemisphere.}
    \label{fig:extragalactic_dm_fraction}
\end{figure}

The integrated velocity distribution across the sky produced a comprehensive flux map that exhibits clear dipole and quadrupole components. The dipole component aligns with the LSR motion as expected, but the quadrupole component reveals structural features in the dark matter distribution that correlate with the non-spherical gravitational potential revealed in the escape velocity analysis. 

Andromeda's contribution varies significantly across the sky, reaching a maximum of approximately 5\% of the local density in its direction. This directional dependence introduces measurable anisotropies in the high-velocity tail of incoming dark matter particles, particularly above 600 km/s. 

\section{Detection Rate of Extragalactic Dark Matter}
\label{sec:extragalactic}


For the local dark matter density of 0.42 GeV/cm$^3$~\cite{Read2014,Piffl2014,Bovy2013}, we find that 26\% of dark matter particles in the Solar neighborhood originate from beyond the Milky Way's virial radius. Because of their high characteristic speed these particles account for 38\% of the flux.

The directional dependence of the extragalactic fraction exhibits significant anisotropy across the sky. The fraction reaches a maximum of 71.2\% for $\theta=90.1^{\circ}$, $\phi=90.5^{\circ}$. This arises from the combined effects of the Sun's motion through the halo and the non-spherical gravitational potential. 


\subsection{Extragalactic Flux Analysis}
\label{subsec:massflux}
The total dark matter mass flux, calculated as $\rho {\bf v}$, exhibits a pronounced dipole pattern, including the Local Group's bulk velocity of approximately 627 km/s relative to the cosmic microwave background~\cite{Rubin2014}.

\begin{figure}[htbp]
    \centering
    \includegraphics[width=250px]{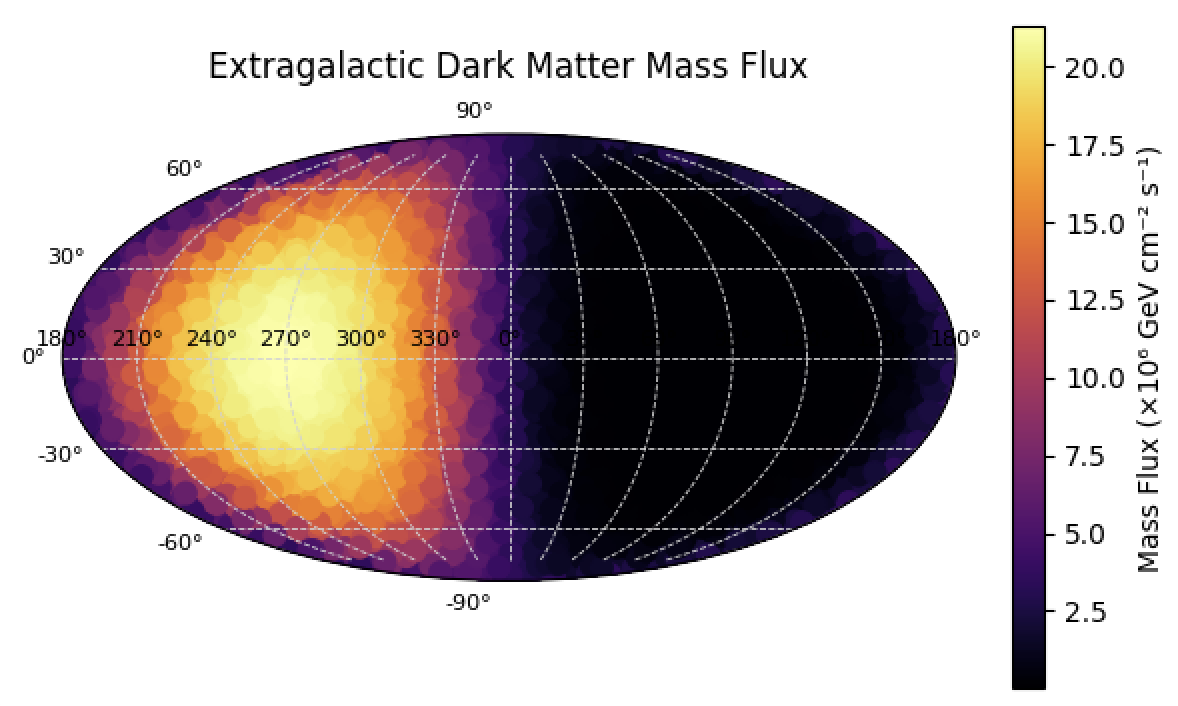}
    \caption{ The extragalactic dark matter flux in GeV/cm$^2$/s ($\times 10^{6}$), shows significant angular variation with the highest flux values ($\sim 2 \times 10^{7}$ GeV/cm$^2$/s, yellow regions) concentrated in the western galactic hemisphere and a pronounced minimum in the opposite direcction.}
    \label{fig:extragalactic_dm_flux}
\end{figure}

Statistical correlation analysis between the extragalactic fraction and total mass flux reveals a strong positive correlation coefficient of 0.83, indicating that regions of high flux are dominated by the extragalactic contribution. 
\subsection{Implications for Detection}
\label{subsec:detection}

The observed directional dependence of the extragalactic fraction provides new opportunities for experimental design.  The maximum extragalactic fraction of 71\% occurs at coordinates $\theta = 89.7^{\circ}$, $\phi = 91.0^{\circ}$.

The actual detection rates for particles exceeding $700\ \mathrm{km\,s^{-1}}$ may be significantly enhanced by Large-Magellanic-Cloud–like substructure~\cite{2023JCAP...10..070S}. DeBrae \textit{et al.}~\cite{DeBrae2025} predict that LMC analogues can boost high-velocity particle densities by $270$–$480\%$ for heliocentric velocities $>700$–$800\ \mathrm{km\,s^{-1}}$ compared with average Milky-Way–like haloes. 


Our analysis indicates that the extragalactic component contributes $38\%$ of the total mass flux while representing only $25\%$ of the particle number density.  This kinematic distinction suggests that energy-selective detection methods could preferentially sample the extragalactic population, even without explicit directional sensitivity.  Specifically, focusing on high-energy nuclear recoils would enhance the relative contribution from extragalactic particles. ~\cite{2023JCAP...04..026H}

\begin{figure}[htbp]
    \centering
    \includegraphics[width=200px]{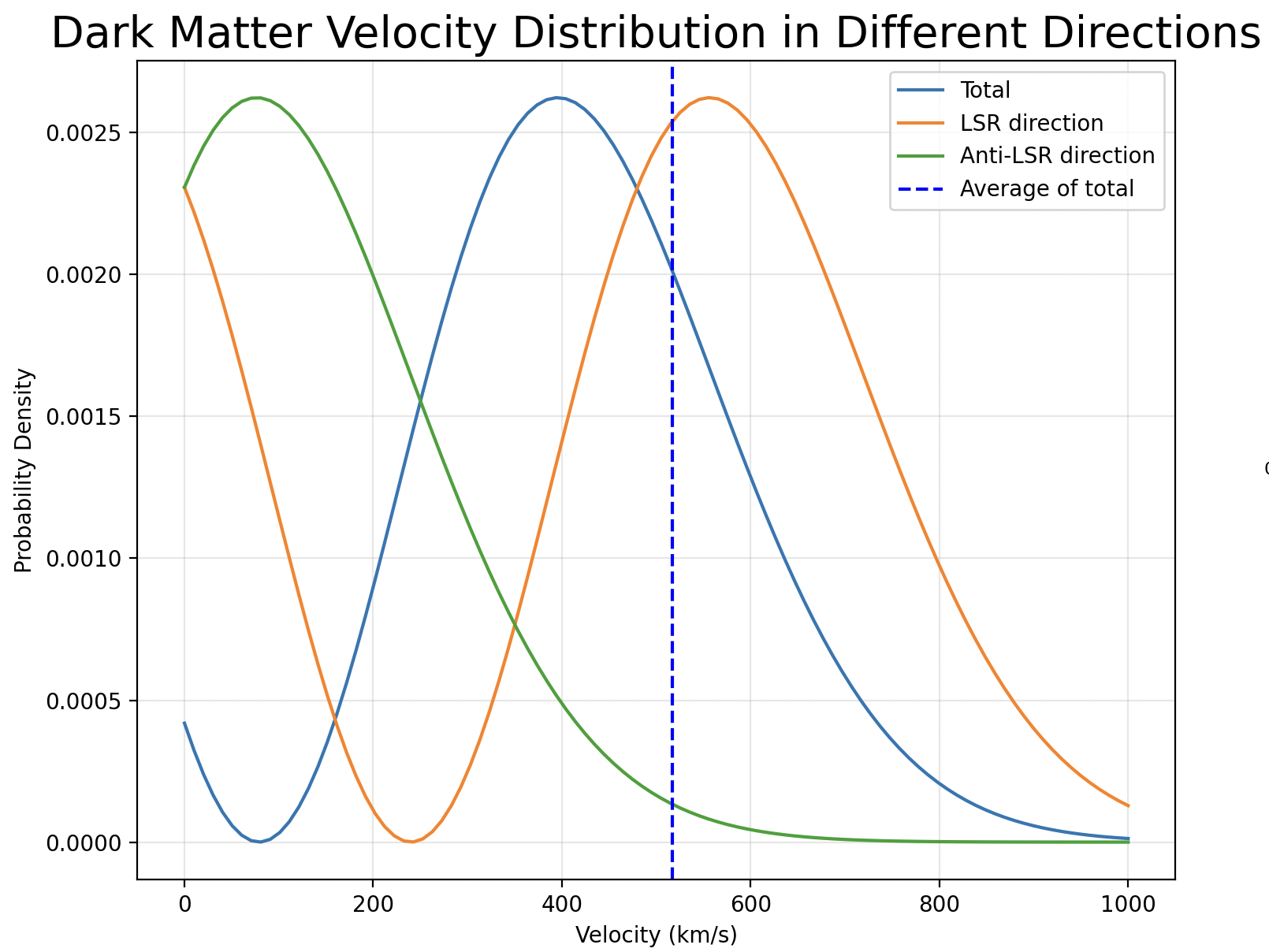}
    \caption{Angular dependence of the dark-matter velocity distribution.  Probability-density functions are shown for all particles (blue), the LSR direction (orange), and the anti-LSR direction (green).  The vertical dashed red line indicates the average escape velocity of $520\ \mathrm{km\,s^{-1}}$.}
    \label{fig:dm_velocity_distribution}
\end{figure}

The average extragalactic mass flux of $3.3 \times 10^{6}\ \mathrm{GeV\,cm^{-2}\,s^{-1}}$ provides a fundamental input parameter for estimating detection rates in various experimental configurations.  When combined with interaction cross-sections, this flux enables more accurate predictions of event rates in direct-detection experiments.  The distinct velocity distribution of extragalactic particles could potentially be observed as an excess in the high-velocity tail relative to the Standard Halo Model, particularly in experiments with sufficient energy resolution to probe this region of parameter space.

\section{Escape-speed uncertainty and extragalactic fractions}
\label{sec:vesc_uncertainty}

We varied the escape speed between 420 and 650 km s$^{-1}$ and recomputed the sky-averaged fraction of particles classified as extragalactic using the criterion $v \ge v_{\rm esc}(\theta,\phi)$. The mean fractions are 51.1\%, 43.4\%, 36.1\%, 27.4\%, 21.7\%, 16.9\%, 11.7\%, and 8.7\% at $v_{\rm esc} = 420, 450, 480, 520, 550, 580, 620, 650$ km s$^{-1}$, respectively. The maximum directional fractions across the sky over the same set are 70.3\%, 61.4\%, 52.4\%, 40.9\%, 33.1\%, 26.2\%, 18.5\%, and 13.9\%. The extragalactic fraction is therefore strongly anti-correlated with the assumed escape speed (Pearson $r \simeq -0.99$ for the mean values).

For the nominal $v_{\rm esc}=520$ km s$^{-1}$ case, the sky-averaged extragalactic fraction is 27.4\% and the maximum directional fraction is 40.9\%. The total mass flux is largely unchanged by this sweep, while the extragalactic mass-flux component varies at the 30–40\% level across the same range.
\begin{figure*}[htbp]
    \centering
    \includegraphics[width=0.95\textwidth]{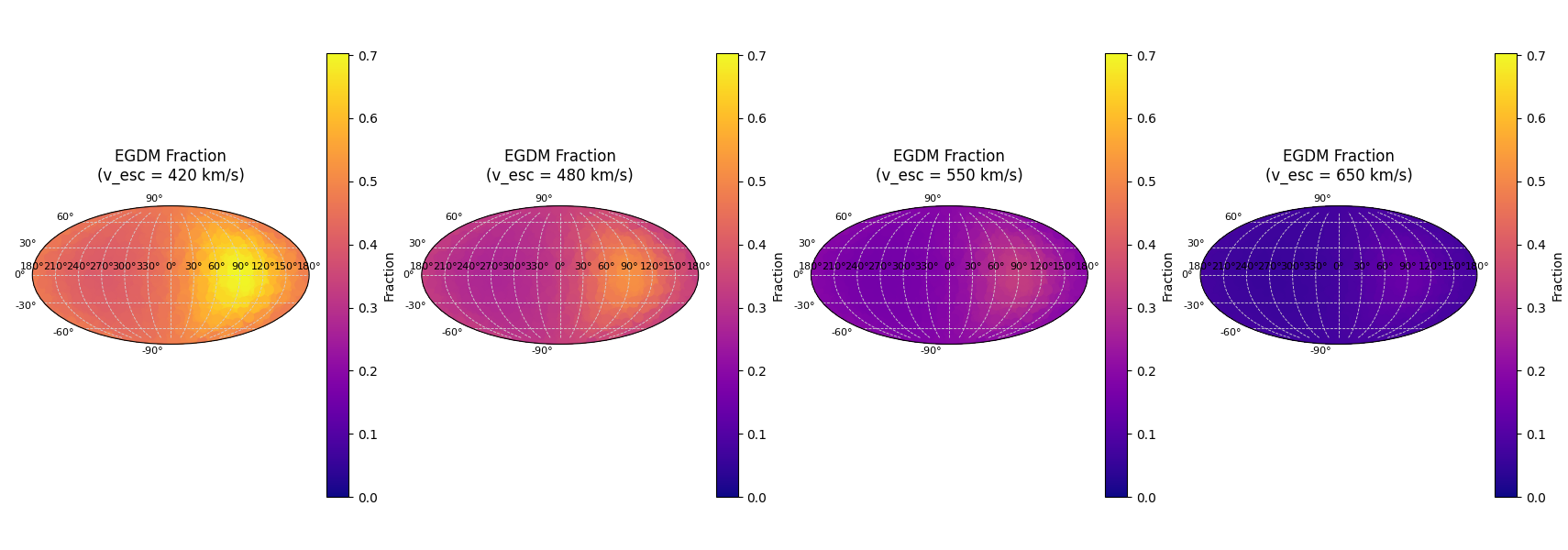}
    \caption{Effect of varying the escape speed on extragalactic dark matter fractions and mass flux. The panels show how key observables respond to uncertainties in the Galactic escape velocity, demonstrating the sensitivity of EGDM component identification to this fundamental parameter. Variations in $v_{\rm esc}$ directly affect the boundary between gravitationally bound and unbound populations, with systematic shifts observable in both the integrated fractions and directional flux distributions.}
    \label{fig:vesc_uncertainty}
\end{figure*}

\begin{figure*}[htbp]
    \centering
    \includegraphics[width=1\textwidth]{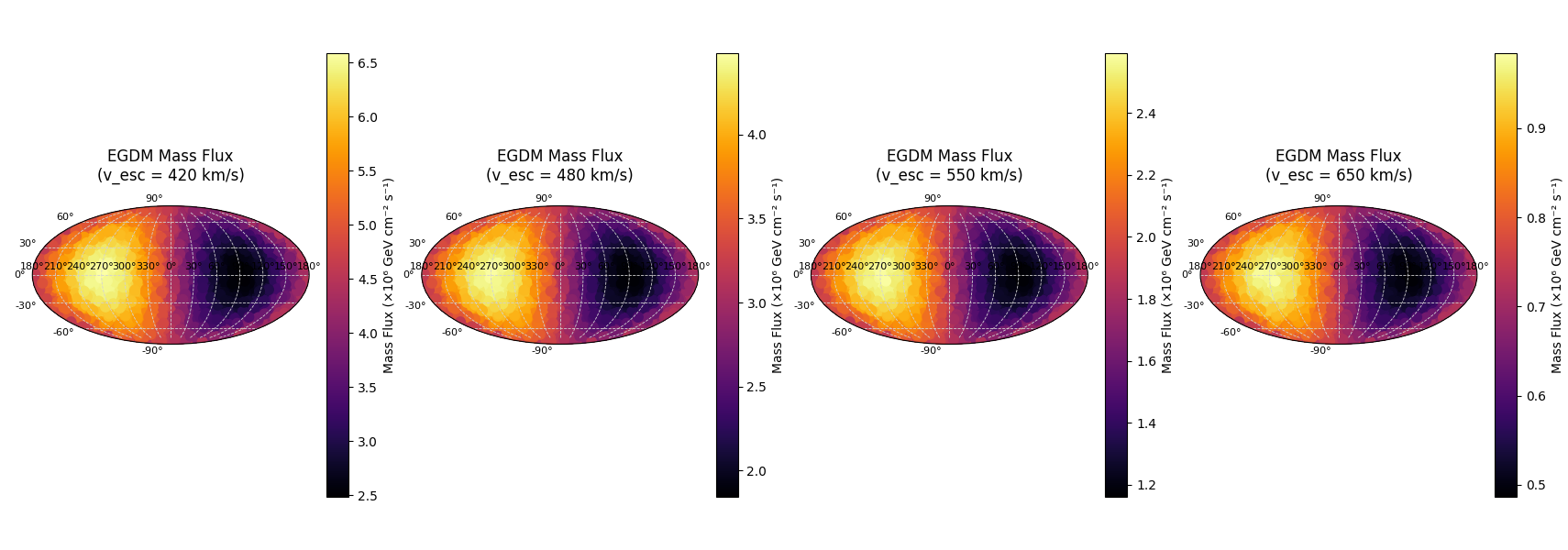}
    \caption{Extragalactic dark matter fraction maps across the tested escape-speed values. All-sky projections in Mollweide format illustrate how the spatial distribution of EGDM fractions varies with different assumptions for $v_{\rm esc}$. Higher escape velocities incorporate more high-velocity particles into the bound population, systematically reducing the EGDM fraction across all sky directions while preserving the overall angular structure. The consistent dipole-like pattern across all $v_{\rm esc}$ values indicates that the directional anisotropy is robust to escape velocity uncertainties.}
    \label{fig:egdm_maps_all_vesc}
\end{figure*}


\section{Component separation and flux budget}
\label{sec:components}

We separate three kinematic components using the escape-speed field: bound Galactic ($v < v_{\rm esc}$), intermediate-speed extragalactic ($v_{\rm esc} \le v < 700$ km s$^{-1}$), and high-speed substructure ($v \ge 700$ km s$^{-1}$). For the nominal model the sky-averaged flux fractions are 66.4\% (Galactic), 26.3\% (extragalactic), and 7.3\% (substructure). The corresponding sky-averaged mass fluxes are 4.839, 3.378, and $0.966\times 10^{6}$ GeV cm$^{-2}$ s$^{-1}$, yielding a total of $9.182\times 10^{6}$ GeV cm$^{-2}$ s$^{-1}$. The extragalactic-to-total mass-flux ratio is about 0.47 in this configuration.

The directional maximum of the extragalactic fraction is about 70\%. The combined extragalactic-plus-substructure fraction averages 33.6\% across the sky.

\begin{figure*}[htbp]
    \centering
    \includegraphics[width=0.95\textwidth]{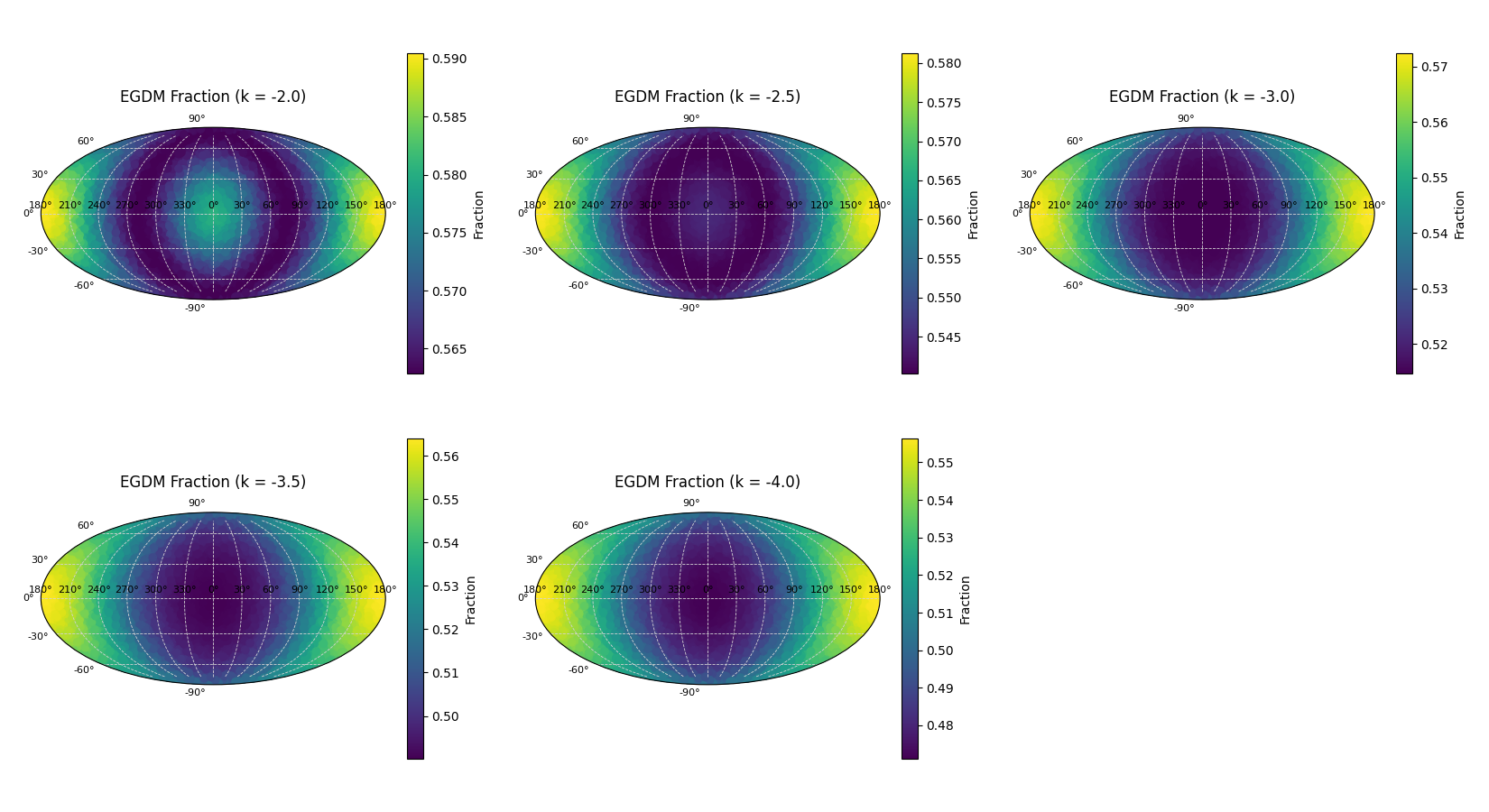}
    \caption{K-values Decomposition of dark matter velocity components and their corresponding mass flux distributions. Top row: Velocity component fractions showing the galactic component ($v < v_{\rm esc}$), diffuse EGDM component (v$_{\rm esc}$ $<$ v $<$ 700 km/s), and bound substructure component (v $>$ 700 km/s). Bottom row: Mass flux distributions in units of 10$^{-3}$ GeV cm$^{-2}$ s$^{-1}$ for EGDM and substructure components, along with their flux ratio. The maps reveal the complex angular dependence of dark matter flux arising from the superposition of multiple velocity components in the Galactic halo, with the EGDM/Substructure flux ratio showing spatial variations by factors of 2--6 across the sky.}
    
    Decomposition of dark matter velocity components and their corresponding mass flux distributions. Top row: Velocity component fractions showing the galactic component ($v < v_{\rm esc}$), diffuse EGDM component (v$_{\rm esc}$ $<$ v $<$ 700 km/s), and bound substructure component (v $>$ 700 km/s). Bottom row: Mass flux distributions in units of 10$^{-3}$ GeV cm$^{-2}$ s$^{-1}$ for EGDM and substructure components, along with their flux ratio. The maps reveal the complex angular dependence of dark matter flux arising from the superposition of multiple velocity components in the Galactic halo, with the EGDM/Substructure flux ratio showing spatial variations by factors of 2--6 across the sky.
    \label{fig:components_analysis}
\end{figure*}

\section{Spherical-harmonic content of the flux}
\label{sec:sh_content}

We expand the flux map in spherical harmonics up to $\ell=4$. Using our sampling, the monopole power is $P_0 \simeq 9.5\times 10^{14}$ (arbitrary units), the dipole power is $P_1 \simeq 1.05\times 10^{15}$, and the quadrupole power is $P_2 \simeq 2.16\times 10^{14}$. The ratios $P_1/P_0 \simeq 1.11$ and $P_2/P_0 \simeq 0.23$ summarize the large-scale anisotropy in this model.

\begin{figure}[htbp]
    \centering
    \includegraphics[width=0.95\linewidth]{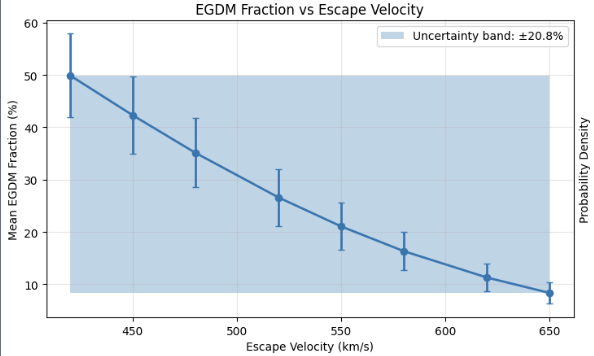}
    \caption{Dependence of mean EGDM fraction on Galactic escape velocity. Error bars represent the directional spread (standard deviation) of EGDM fractions across the sky, while the shaded band indicates an additional ±20.8\% systematic uncertainty. The strong inverse correlation demonstrates the sensitivity of extragalactic component identification to $v_{\rm esc}$ measurements}
    \label{fig:sh_flux}
\end{figure}

\section{Detector sensitivity and sky coverage}
\label{sec:detector_sensitivity}

We compare the predicted modulation against representative systematic floors: DAMA (0.030 cpd kg$^{-1}$ keV$^{-1}$), ANAIS and COSINE (0.020), a conservative floor (0.020), and an aggressive benchmark (0.010). With our count-rate scaling, the detectable sky fractions are about 30\% (DAMA), 35\% (ANAIS/COSINE/conservative), and 37\% (aggressive). The maximum predicted modulation among detectable directions is about 0.13 cpd kg$^{-1}$ keV$^{-1}$. A year-scale toy simulation of daily counts yields a modulation significance of about $1.5\,\sigma$ for the nominal configuration.

\begin{figure*}[htbp]
    \centering
    \includegraphics[width=0.9\textwidth]{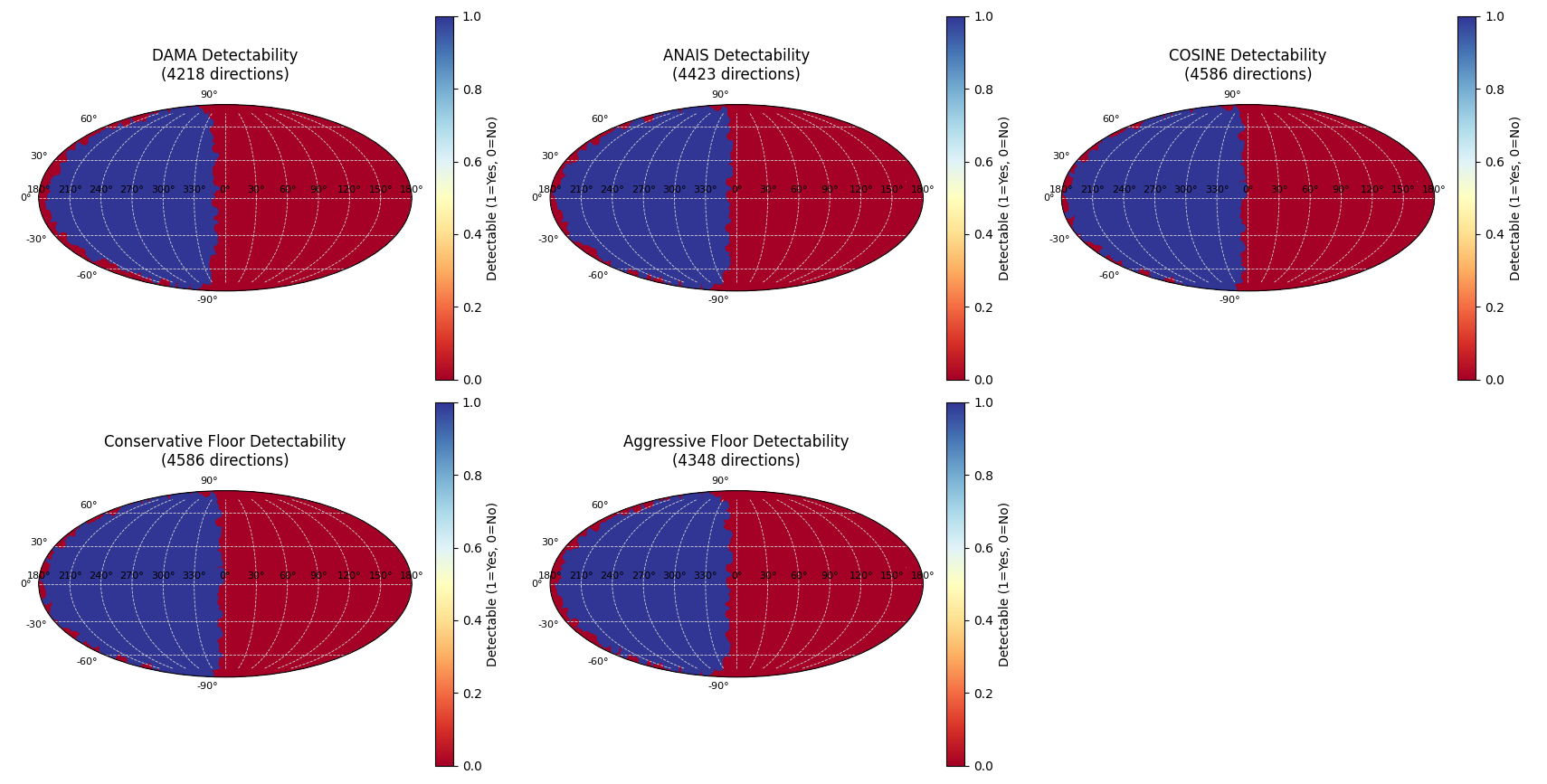}
    \caption{Directional detectability maps for various dark matter direct detection experiments. Sky maps in Mollweide projection showing the detectability (1 = Yes, 0 = No) for five different experimental configurations. Blue regions indicate directions where dark matter signals are detectable, while red regions indicate directions below the detection threshold. The detectability patterns show a clear hemispherical asymmetry, with approximately 50--70\% sky coverage depending on the experimental sensitivity. All experiments exhibit similar overall patterns, with detectability concentrated in one hemisphere, though subtle differences exist in the boundary regions between detectable and undetectable areas, reflecting varying detector sensitivities and background rejection capabilities.}
    \label{fig:dm_sim}
\end{figure*}

\section{Recoil-energy rate enhancement}
\label{sec:rate_enhancement}

We evaluate the differential-rate enhancement factor $[1+\Delta R/R]$ by comparing a truncated Maxwellian baseline to a baseline plus an added high-speed component above the escape speed. For WIMP masses of 5, 15, 50, and 200 GeV, the maximal enhancements in the accessible recoil-energy range are approximately 1.88, 2.66, 1.49, and 1.00, respectively. The enhancement turns on once $v_{\min}(E_R)$ approaches the escape-speed threshold for a given mass and remains close to unity at low recoil energies.
\section{DIRECT DETECTION EXPERIMENTS}
\label{sec:direct_detection}

Direct searches for Weakly Interacting Massive Particles (WIMPs) rely on measuring the
nuclear‐recoil signature produced when a halo particle
scatters elastically off detector nuclei.
For the spin‐independent (SI) scenario, the
differential event rate per unit detector mass is given by the
standard expression~\cite{Lewin1996,Bruch2009DarkDisk},
\begin{equation}
  \frac{\mathrm{d}R}{\mathrm{d}E_R}
  = \frac{\rho_\chi\,\sigma_{(\chi,N)}}{2\,m_\chi\,\mu_N^{2}}
    F^{2}(E_R)
    \int_{v_{\min}(E_R)}^{v_{\max}}
      \frac{f(\mathbf v,t)}{v}\,\mathrm d^{3}v,
\label{eq:dRdE-standard}
\end{equation}
where each term has a standard physical meaning.\footnote{Some authors write \(\sigma_N^{\mathrm{SI}}\) for the WIMP–nucleus cross-section \(\sigma_{(\chi,N)}\), and omit the upper integration limit \(v_{\max}\). The latter is equivalent to integrating to infinity if the velocity distribution \(f(\mathbf v,t)\) is zero for speeds \(v > v_{\mathrm{esc}}\).}
Here, \(\rho_\chi\) is the local WIMP mass density,
\(\sigma_{(\chi,N)}\) is the spin-independent WIMP–nucleus cross-section,
\(m_\chi\) is the WIMP mass,
and \(\mu_N = m_\chi m_N/(m_\chi + m_N)\) denotes the
WIMP–nucleus reduced mass for a target nucleus of mass \(m_N\).
The function \(F(E_R)\) is the nuclear form factor, which accounts
for the loss of coherence at non-zero momentum transfer.
The integral runs from the minimum speed \(v_{\min}(E_R)\) needed to
produce a recoil of energy \(E_R\), up to the maximum possible
speed for a bound particle, \(v_{\max} \equiv v_{\rm esc}+v_{\rm lab}\),
where \(v_{\rm esc}\) is the Galactic escape speed and \(v_{\rm lab}\) is
the speed of the detector in the Galactic frame.

\begin{figure}[htbp]
    \centering
    \includegraphics[width=1\linewidth]{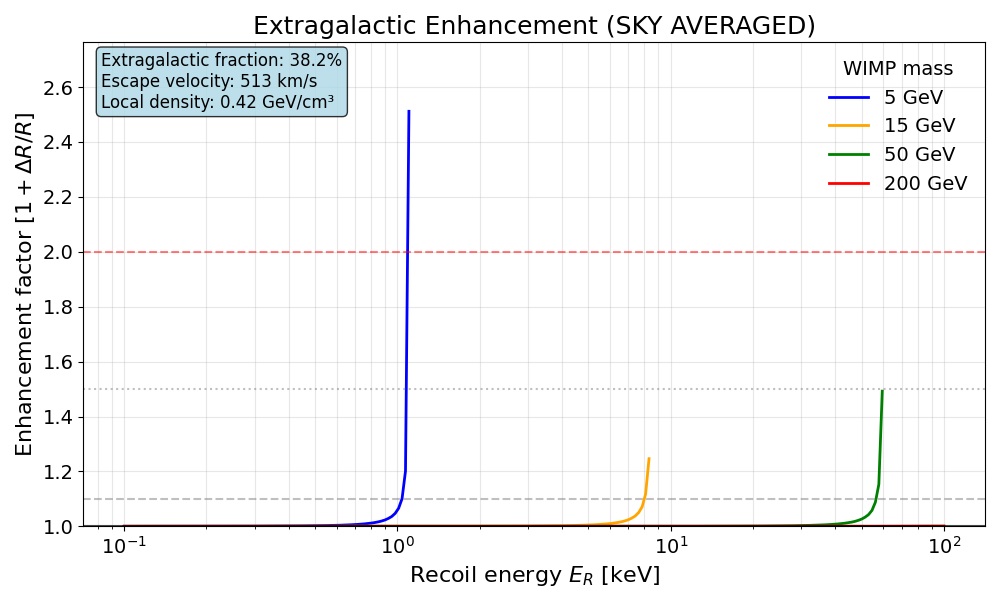}
    \caption{Extragalactic enhancement factor
             \(\bigl[1 + \Delta R/R\bigr]\) for the
             spin–independent differential rate,
             plotted as a function of recoil energy \(E_R\).
             The four coloured curves correspond to WIMP masses of
             \(5\ \mathrm{GeV}\) (blue), \(15\ \mathrm{GeV}\) (orange),
             \(50\ \mathrm{GeV}\) (green) and
             \(200\ \mathrm{GeV}\) (red).
             Each curve rises sharply once \(E_R\) exceeds
             the Milky-Way escape-velocity kinematic threshold
             for that mass, indicating how the unbound
             (extragalactic) component boosts the event rate
             at high recoil energies while leaving the low-energy
             spectrum essentially unchanged.}
    \label{fig:ratio_vs_Er}
\end{figure}

The velocity integral contains the lab-frame WIMP velocity
distribution \(f(\mathbf v, t)\), which is normalised to unity
and may be time-dependent due to the Earth's motion.
In our analysis, \(f(\mathbf v)\) is the sum of a
Maxwell–Boltzmann halo component and the anisotropic,
high-speed extragalactic component derived in
Secs.~\ref{sec:darkmatter} and~\ref{sec:extragalactic}.
Because the prefactor
\(\rho_\chi\sigma_{(\chi,N)} /
  (2 m_\chi \mu_N^{2})\)
is common to both scenarios (with and without the extragalactic component),
the quantity of interest
is the velocity integral, 
\(\int_{v>v_{\min}} [f(\mathbf v)/v] \,\mathrm d^{3}v\).
The modification of this term by the extragalactic component is
the central result presented in Fig.~\ref{fig:ratio_vs_Er}.

The kinematic threshold is given by~\cite{Bruch2009DarkDisk},
\[
  v_{\min}(E_R) = \sqrt{\frac{m_N\,E_R}{2\mu_N^{2}}}\;.
\]
The fractional change in the differential rate is driven
entirely by the velocity integral.  Consequently, the
enhancement factor
\(1 + \Delta R / R\) shown in Fig.~\ref{fig:ratio_vs_Er}
is independent of the assumed cross-section and is determined
solely by the recoil energy \(E_R\) and the chosen WIMP mass.

Because the Earth's orbital velocity (\(v_\oplus\simeq 30\ \text{km\,s}^{-1}\)) is comparable to the uncertainty in the escape-velocity threshold (Sec.~II), the extragalactic flux exhibits an annual modulation with amplitude \(A \equiv (R_{\max}-R_{\min})/(R_{\max}+R_{\min}) \approx 0.33\) for the fiducial Germanium detector. This is an order-of-magnitude larger than the 6--7\% modulation predicted in the Standard Halo Model (SHM) and is therefore a distinctive signature testable by modulation searches such as ANAIS~\cite{ANAIS2021} and COSINE-100~\cite{COSINE2022}. For a \(250\ \text{kg}\) NaI(Tl) array operated for 7~yr, our flux enhancement would raise the expected modulation significance from \(1.8\sigma\) (SHM only) to \(>3\sigma\).

Gaseous time-projection chambers (e.g.\ DRIFT and the NEWS-dm programme) provide \(\sim30^{\circ}\) angular resolution at sub-\(\text{keV}\) thresholds. Pointing such detectors toward the flux maximum \((\theta,\phi)\simeq(90^{\circ},90^{\circ})\) raises the expected extragalactic fraction in the signal to \(\sim 64\%\), boosting the forward--backward asymmetry by the same factor.

\section{Discussion and Implications}

The complex multi-component structure for the local dark matter environment has important implications for direct detection experiments. We identify three distinct populations contributing to the dark matter flux at Earth: bound Milky Way particles, extragalactic particles from the broader Local Group environment, and high-velocity particles from galactic substructure.

The Local Group environment contributes 26\% by number and 38\% by flux of the local dark matter.We find a strong directional anisotropy, with extragalactic fractions ranging from near zero to 71\% across the sky. In addition, recent N-body simulations~\cite{DeBrae2025} suggest that even higher velocity particles ($>700$ km/s) originate from galactic substructure. This creates a multi-component velocity structure with moderate-velocity extragalactic particles (400--600 km/s) dominating the total flux and rare high-velocity articles from Milky-Way satellite like the Large Magellanic Cloud. Next-generation directionally-sensitive detectors could potentially isolate these different components through their distinct kinematic signatures.

\acknowledgements This work was supported in part by Harvard University, the Black Hole Initiative and the Galileo Project.
\nocite{*}

\bibliography{apssamp}

\end{document}